\documentclass[prl, reprint, nofootinbib, showpacs,preprintnumbers,amsmath,amssymb]{revtex4}
\usepackage{varioref,exscale,latexsym,amsmath,amssymb}
\usepackage{graphicx}

\usepackage{graphicx}
\usepackage{slashed}
\usepackage{dcolumn}
\usepackage{bm}
\usepackage{hyperref}
\usepackage{color}
\usepackage{xcolor}
\usepackage[export]{adjustbox}

\newcommand{\beq}{\begin{equation}}
\newcommand{\eeq}{\end{equation}}
\newcommand{\bea}{\begin{eqnarray}}
\newcommand{\eea}{\end{eqnarray}}
\newcommand{\nn}{\nonumber}

\def\lsi{\raise0.3ex\hbox{$<$\kern-0.75em\raise-1.1ex\hbox{$\sim$}}}
\def\gsi{\raise0.3ex\hbox{$>$\kern-0.75em\raise-1.1ex\hbox{$\sim$}}}

\def\beq{\begin{equation}}

\def\eeq{\end{equation}}

\def\cN{\mathcal{N}}

\def\cM{\mathcal{M}}

\begin{document}
\preprint{ACFI-T21-06}

\title{{\bf Causality and gravity}}

\medskip\

\medskip\

\author{John F. Donoghue${}^{1}$}
\email{donoghue@physics.umass.edu}
\author{Gabriel Menezes${}^{2}$}
\email{gabrielmenezes@ufrrj.br}
\affiliation{
${}^1$Department of Physics,
University of Massachusetts,
Amherst, MA  01003, USA\\
${}^2$Departamento de F\'{i}sica, Universidade Federal Rural do Rio de Janeiro, 23897-000, Serop\'{e}dica, RJ, Brazil}

\begin{abstract}
We show how uncertainty in the causal structure of field theory is essentially inevitable when one includes quantum gravity. This includes the fact that lightcones are ill-defined in such a theory. This effect is small in the effective field theory regime, where it it independent of the UV completion of the theory, but grows with energy and represents an unknown uncertainty for a generic UV completion.  We include  details of the causality uncertainty which arises in a particular UV completion, i.e.  quadratic gravity. We describe how the mechanisms uncovered in the effective field theory treatment, and some of those in quadratic gravity, could be common features of quantum gravity.
\end{abstract}
\maketitle

\section{Introduction}

In our classical world, causality is taken to mean that there is no effect before its cause. Technically, we therefore use retarded propagators, which vanish for propagation backwards-in-time, to describe classical physics. Quantum physics is a bit different. Using Feynman boundary conditions, propagators do have a backwards-in-time propagation for so-called negative energy modes\footnote{These are antiparticle modes for fermions or charged particles, but gravitons are their own antiparticle so the modes are best categorized as negative energy.}. Indeed this is required \cite{Weinberg:1995mt} in order to obtain a more general definition of causality - that field operators commute for spacelike separations \cite{GellMann:1954db}. When combined with a definition of past and future \cite{Donoghue:2019ecz, Donoghue:2020mdd}, this guarantees that events are only influenced by features within their past lightcone. The commutating of field operators is easy to prove for free fields, although harder to prove for interacting theories. The result of this definition of causality is reflected in the analyticity properties of scattering amplitudes, including relations between the Euclidean and Minkowski amplitudes.

In gravitational quantum field theory, causality is a more difficult concept because spacetime is not the fixed rigid structure of Minkowski QFTs. Even the idea of a particle becomes difficult to define globally. Analyticity properties no longer hold as Euclidean and Lorentzian versions of non-trivial dynamical spacetimes are different. At a handwaving level, we would also expect that fluctuations in the structure of spacetime would  lead to violations or at least uncertainties in the causal of field theories of gravity. This heuristic argument is part of our motivation to see what explicit calculations say about causality in the presence of quantum gravity effects. That is the goal of this paper.

Because of the limitations of understanding of quantum field theory in general curved spacetimes, we are limited somewhat to perturbative calculations relatively close to flat space. For these situations, the effective field theory of gravity is useful, and we can study extended quantum field theories which attempt to go beyond the effective field theory. However, even in this limited setting we can see that there will be an inherent uncertainty in our usual concepts of causality in a gravitational setting. Perhaps the effects will be more than just uncertainty, leading to actual violations at short distance scales. We discuss the calculations which lead to these conclusions below.

\section{Overview of causality in quantum field theory}

For a free field, it is easy to use canonical quantization to calculate the commutator of two fields
\beq
[\phi(x), \phi(y) ] = \int\frac{d^3 k}{(2\pi)3 2E_k} \left[e^{-ik\cdot(x-y)} - e^{+ik\cdot(x-y)} \right]  \  \ .
\eeq
Because the integration $\int{d^3 k}/{(2\pi)3 2E_k}$ is Lorentz invariant, when $(x-y)^0<0$ (i.e. spacelike in our metric) we can transform this to a frame where $(x_0-y_0)=0 $. There the equal time commutator vanishes, as can be readily seen by changing the spatial vector from $\vec{k}$ to $- \vec{k}$ in the second term.

However this does not imply that all propagation is forward in time. The Feynman propagator
\beq
D_F(x-y) = \langle 0 |T\phi(x)\phi(y) |0\rangle = \int \frac{d^{4} k}{(2 \pi)^{4}} e^{-i k \cdot(x-y)} \frac{ i}{k^{2}-m^{2}+ i \epsilon}
\eeq
can be evaluated for $x_0 -y_0 < 0$, with the result
\beq
D_{F}(x-y) = \int \frac{ d^3q}{(2\pi)^3 2E_q} e^{i(E_qt - \vec{q}\cdot\vec{x})} ~~~~{\rm when} ~~~~~x_0 -y_0 < 0 \ \ .
\eeq

We have discussed elsewhere in more detail how the factor of $i\epsilon$ determines the arrow of causality, and how it emerges from the factors of $i$ in the quantization procedure\cite{Donoghue:2019ecz, Donoghue:2020mdd}. In its simplest and briefest form, it can be seen in the path integral quantization
\bea\label{usual}
Z[J] &=& \int [d\phi ] e^{+ i S(\phi, J)}
\nn\\
&=& \int [d\phi ] e^{+ i \int d^4 x [\frac12 (\partial_\mu \phi \partial^\mu \phi - m^2 \phi^2 )+ J\phi ]}  \ \ .
\eea
using $e^{iS}$ rather than $e^{-iS}$. In path integral quantization, this is the only place where $i$ enters the theory. The operation of time reversal $T$ is antiunitary, changing the sign of $i$ and reversing the direction of causality.

In a gravitational theory, there is no global timelike direction in a generic background. Nor are there global Lorentz boosts, such as employed in the proof of the vanishing of the commutator. Nevertheless, classically there are well-defined light cones as all massless particles follow a common geodesic. Attempts to construct propagators on a classical background hypothesize that the free field commutators will still vanish outside of the backwards classical lightcone. While this is plausible, it becomes a daunting requirement if the spacetime is itself quantum mechanically dynamical.

At low energy, the quantum effects of gravity can be treated using effective field theory techniques\cite{Donoghue:1994dn, Donoghue:2017pgk}. These are universal in the sense that they are independent of any particular ultraviolet (UV) completion, as they depend only on the degrees of freedom and the interactions at low energy. At very low energies, all effects which have been calculated in the effective field theory are small, as they are suppressed by the weakness of gravity at low energy. However, these effects grow in magnitude as the energy increases and become of order unity at the Planck scale. This signals the need to have a modification of General Relativity when one reaches the Planck scale. However, the effective field calculations can still be instructive as the demonstrate the effect of mechanisms, such as spacetime nonlocality, which are expected to survive in all UV completions.

There is a particular UV completion of gravity which we find useful in discussing causality. This is quadratic gravity \cite{Stelle:1976gc, Salvio:2018crh, Strumia, Donoghue:2018izj, Holdom, unitarity} with the action
\beq
S_{\textrm{quad}} = \int d^4x \sqrt{-g}
\left[\frac{2}{\kappa^2}\sqrt{-g} R+\frac{1}{6 f_0^2} R^2 - \frac{1}{2\xi^2} C_{\mu\nu\alpha\beta}C^{\mu\nu\alpha\beta}\right]
\eeq
where $\kappa^2 =32\pi G$ and $f_0,~\xi$ are coupling constants. $C_{\mu\nu\alpha\beta}$ is the Weyl tensor, where
\beq \label{weylidentity}
- \frac{1}{2\xi^2} C_{\mu\nu\alpha\beta}C^{\mu\nu\alpha\beta} =  -\frac{1}{\xi^2}\left[\left(R_{\mu\nu}R^{\mu\nu}
- \frac13 R^2\right) +\frac{1}{2} G\right]  \ \ .
\eeq
This is a renormalizable field theory. For our purposes one of the key features of this UV completion is that it connects naturally to the effective field theory. Often UV theories involve different degrees of freedom than those involved with the effective field theory. A classic example is the quarks and gluons of QCD vs the pions of the low energy chiral effective field theory. With gravity the range of hypothesised fundamental degrees of freedom is even greater. However, in quadratic gravity the metric is the fundamental degree of freedom at all scales. Moreover, the effective field theory at low energy is well defined from the fundamental theory, a feature that almost no other UV completions can provide.

One of the special features of quadratic gravity is an unstable high mass ghost in the spin-two channel. To see this, we note that the higher curvature terms involve four derivatives and lead to quartic terms in the propagator. In the spin-two channel, including the vacuum polarization diagram of the matter fields, this leads to
\begin{eqnarray}
iD_{\mu\nu\alpha\beta} &=& i{\cal P}^{(2)}_{\mu\nu\alpha\beta} D_2(q)
\nonumber \\
D^{-1}_2(q) &=& {q^2+i\epsilon}- \frac{\kappa^2 q^4}{2\xi^2(\mu)}
- \frac{\kappa^2 q^4 N_{\textrm{eff}}}{640\pi^2} \ln \left(\frac{-q^2-i\epsilon}{\mu^2}\right)
\end{eqnarray}
Here, ${\cal P}^{(2)}_{\mu\nu\alpha\beta}$ is the spin-two projector
\beq
{\cal P}^{(2)}_{\mu\nu\rho\sigma} = \frac{1}{2}(\theta_{\mu\rho}\theta_{\nu\sigma}
+ \theta_{\mu\sigma}\theta_{\nu\rho}) - \frac{1}{3} \theta_{\mu\nu}\theta_{\rho\sigma}  \ \ ,
\eeq
and $N_{eff}$ are effective number of light degrees of freedom\footnote{There are also some logarithmic contributions which do not contribute to imaginary parts in the propagator \cite{Donoghue:2018izj} which are not displayed here.}. Besides the usual graviton pole at $q^2=0$ there is a high mass pole and near that pole the propagator has the form
\beq
iD_2 (q) \sim \frac{-i}{q^2-m^2 -i\gamma}
\eeq
where $\gamma >0 $ follows from the factor of $i\epsilon$ in the logarithm.

As far as causality is concerned this has two relevant features - the sign in the numerator and the sign of the imaginary part in the denominator. Both of these are changes of $i$ to $-i$ and indicate that this corresponds to time-reversed propagation. More detailed analysis of the propagators confirms this \cite{Donoghue:2019ecz}. We have called this resonance a {\em Merlin mode} because it propagates backwards in time compared to the usual modes.

Quadratic gravity is of course only one of the possible UV completions of gravity. However because it forms a field theory, and because we can easily match on to the effective field theory, we find it a useful example for our discussion. Moreover, as it is in some ways the most conservative UV completion, it may  be relevant to Planck scale phenomenology.

\section{The lightcone is not a quantum gravity concept}

Many discussions of quantum gravity use concepts which are carried over from classical gravity, such as lightcones, geodesics and Penrose diagrams. However, these are not well defined in quantum gravity. A necessary consequence is that causality becomes a fuzzy concept in gravitational backgrounds, even if they are asymptotically flat.

{ The problems can in principle arise due to higher order operators in the gravitational Lagrangian. For example, the operator $R_{\mu\nu}R^{\mu\nu} $ will appear in the gravitational Lagrangian in order to absorb divergences which occur at one loop order. However, this operator will not influence graviton lightcones, because it vanishes for on-shell gravitons. In the next section, we discuss the effect of this operator when it is treated off-shell. The result is that in the effective field theory limit this is a small correction to the graviton propagator, but it grows as the energy scale approaches the Planck scale. Ultimately it leads to causality violation, which is fortunately exponentially suppressed even at high energy. There is also the possibility of higher order corrections to the gravitational couplings. In Ref \cite{Camanho:2016}, the effects of the three-point couplings on causality were explored. These were power suppressed, and are small in the region where the effective field theory treatment is valid.  Interestingly, even when extrapolating to small distances beyond the effective field theory limit, it was found that for all the considered interactions the causality violating effects vanished in four dimensions, except for those which were time-reversal violating. The authors of \cite{Camanho:2016} phrased the latter condition as parity-violation. However, since the interactions considered are even under charge conjugation, by the CPT theorem this condition is the same as time-reversal violation. Such effects could be potentially important for causality, but are in some sense optional and can be excised by imposing time-reversal symmetry on the gravitational interaction. There have also been some suggestions that spin effects violate the equivalence principle and lead to non-unique trajectories. This seems to have been resolved \cite{Silenko:2004ad} in favor of the classical equivalence principle behavior. }

{However, quantum corrections to lightcones are not optional. They can be calculated in the effective field theory limit of gravity, as described below. These are small at low energy, but grow to be of order unity by the Planck scale. While the specifics of this growth becomes less reliable as one approaches the Planck scale, as a more complete theory may supplant the effective field theory, the underlying reasons behind these effects are unlikely to disappear because they are general features of quantum corrections. We turn to these quantum corrections next. }

Classically any massless particle defines the lightcone. Since the speed of propagation is the same for all massless particles, these lightcones would be identical for all forms of massless particles. However in a gravitational background this is no longer true. As an example the gravitational bending angle of a massless particle around a heavy mass M has been calculated and is found to have the form \cite{Bjerrum-Bohr:2016hpa, Bjerrum-Bohr:2014zsa, Bai:2016ivl, Chi:2019owc}
\begin{equation}
\theta \simeq \frac{4 G_{N} M}{b}+\frac{15}{4} \frac{G_{N}^{2} M^{2} \pi}{b^{2}}+\left(8 c^{S}+9-48 \log \frac{b}{2 b_{0}}\right) \frac{\hbar G_{N}^{2} M}{\pi b^{3}}+\ldots
\end{equation}
with $c^{S}$ being a constant which depends on the spin of the scattered particle. It is found that $c^{S} = 371/120,~ 113/120, ~-29/8    $ for a massless scalar, photon or graviton, respectively. Here $b$ is the impact parameter and $\log b_0$ is an infrared logarithm which we need not be concerned with here. This result has been calculated in the eikonal limit. This involves large impact parameter, and is the limit where the effective field theory is most applicable. Of course, quantum scattering leads to a range of bending angles and this bending angle formula has been defined by using the maximum of the eikonal phase.

There are several implications which can be immediately seen. This phenomenon cannot be described by geodesic motion as different massless species respond differently. It is then not equivalent to a quantum modification to the metric. The trajectory of massless particles is used in flat space to define the lightcone. That cannot be done in this background again due to the species dependence. Moreover, in the details of the calculation we can see that the quantum evolution samples the gravitational field over many points in space, not just along a local geodesic. In loop diagrams, the massless propagators are not localized in spacetime, but propagate over large distances. This then is sensitive to the gravitational field over these distances. When averaged over all directions, this converts the leading $1/b$ dependence into the $1/b^2$ classical correction and the $1/b^3$ quantum correction. The nonlocality is manifest in non-analytic terms in the amplitudes. These cannot be Taylor expanded in the momentum. Taylor expansions of analytic terms can be written as local operators, with the higher momentum pieces being equivalent to derivatives of delta functions. The non-analytic and non-universal behaviors appear even in the gravitational couplings. This can be seen from the effects of long distance graviton propagators, as calculated in \cite{Bjerrum-Bohr:2016hpa, Bjerrum-Bohr:2014zsa}, with the amplitude,
\begin{eqnarray} \label{e:ample}
&&    i
 \cM_X
 \simeq{\cN^{X}\over \hbar}\,  (M\omega)^2\cr
&&\times
 \Big[-{\kappa^2\over \boldsymbol q^2}+\kappa^4 {15\over
  512}{ M\over |\boldsymbol q|}
+\hbar \kappa^4
 {15\over
  512\pi^2}\,\log\left(\boldsymbol q^2\over M^2\right)
-\hbar\kappa^4\,{ bu^{S} \over(8\pi)^2} \, \log\left(\boldsymbol q^2\over \mu^2\right)
\\
\nonumber &&+ \hbar\kappa^4 {3\over128\pi^2}\, \log^2\left(\boldsymbol q^2\over \mu^2 \right)
-\kappa^4 \,  {M\omega\over 8\pi} {i\over \boldsymbol q^2}\log\left(\boldsymbol q^2\over
  M^2\right)
\Big]\,.
\label{result1}\end{eqnarray}
Here $\cN^{X}$ is an overall normalization, $\omega$ is the energy and $q^2$ is the four-momentum transfer squared. The point to notice here is that the logarithms, and squares of logarithms, signal the non-locality in spacetime. Logarithms cannot be Taylor expanded around $q^2 =0$ and therefore their Fourier transform cannot be written in a derivative expansion of local terms. In addition, the coefficient $ bu^{S} $ differs for different species, much like the coefficient $c^{S}$ in the formula for the bending angle given above. This was calculated using a unitarity-based method, and the fact that this occurs in the graviton cuts is another indication that it is really quantum gravity participating in this non-locality and non-universality. Overall, this non-locality implies that the motion is not purely a geodesic.

Of course the eikonal amplitude is derived in a limit of a weak field and large impact parameter.

The actual quantum modification for motion aroung the Sun is tiny, and for all effective purposes we can continue to use lightcones in situations where the quantum graviy effects are small. The effective field theory calculations are nevertheless valuable because they are universal.  Moreover one can argue that the modifications found here are generic in that even at higher energies they will happen in any quantum theory of gravity. There are anomalous corrections because quantum effects are not local. The propagation of massless particles such as the graviton can occur over large distances and the particle in question will sample the gravitational field not only along the classical trajectory but also in the neighborhood of that trajectory. With massless particles, that neighborhood is not well localized. When the gravitational field is more significant, we expect that these non-local effects will be even larger. Therefore, we expect that the lessons learned here will carry over to stronger and more complicated gravitational backgrounds. The propagation of massless particles in loops will always make the interaction nonlocal - sampling the background field over a range of positions. The effective field theory calculation has identified a generic mechanism. The trajectory (however it is defined) will not be described by the classical geodesic. Like the eikonal amplitude, we should expect that the effects are different for different species. This rules out the possibility of defining an effective geodesic.  Moreover the wave nature of the massless fields becomes more important at strong coupling and it is not possible to define an effective ``bending angle'' as was done in the simpler eikonal approximation.

These simple considerations tell us that many of our familiar classical concepts do not carry over to quantum gravitational physics. For causality, it implies that we lose our standard definitions of causal behavior, even if the asymptotic observers are in regions which are flat enough to apply Minkowski definitions locally~\footnote{We remark that the discussion of lightcone fluctuations here differ from those developed by Ford and collaborators, see for instance Refs.~\cite{Ford:95,Ford:96,Ford:99}. In such analyses, even though one can verify  ``faster than light" signals, there is no causality uncertainty since the system is no longer Lorentz invariant -- the graviton state specifies a preferred frame of reference. Another low energy treatment of the causal uncertainty can be found in \cite{Padmanabhan:1986hd}.}.

\section{Departure from low energy - time delay in phase shifts}

One might be concerned that causality uncertainties from high energy effects could be manifest at low energy and spoil the causal properties of the effective field theory. One low energy manifestation is in the signs of coefficients of the local terms in an effective Lagrangian. This was first noted in non-gravitational effective Lagrangians \cite{Adams:06}. Some of the higher order coefficients must be positive (in a given convention) in order to emerge from a causal UV complete theory. There has been an effort to extend these constraints to gravitational theories also. Here the arguments are not as clear as they invoke classical gravitational concepts such a lightcones, and so could be viewed with caution. However, we will show below that as one departs from low energy there are effects which also traditionally signal causality violation and which do depend on the sign of the coefficients, without directly invoking the lightcone.

If there is a coefficient with the ``wrong'' sign in the low energy effective Lagrangian, does this upset causality in the effective field theory in the region where the effective field theory is valid? A recent analysis \cite{Tolley20} shows that it does not. The point is that in the low energy limit one deals with fields with large wavelengths. The higher order operators in the theory are small perturbations in the effective field theory limit. While there can be time advances and superluminality generated by these operators, the effects are small enough to be unresolvable when working with large wavelength fields which are appropriate for the effective field theory.

However, as one proceeds to higher energy, the effects of the higher order operators will become more apparent. Their effects can become relevant as one approaches the limits to the validity of the effective field theory, even if new degrees of freedom are not yet dynamically active. This can be seen in standard effective field theories, such as chiral perturbation theory. For example in the $J=0$ channel of $\pi\pi$ scattering, the amplitudes approach the unitarity bounds well below the GeV scales where the quarks and gluons become active. The theory still respects unitarity in this limit \cite{Aydemir:2012nz}, but becomes strongly coupled. If the chiral Lagrangian had ``wrong'' sign coeficients (it does not, because QCD is causal) they could show up in scattering phases in this transition region.

We present an example from quadratic gravity which has this behavior. In quadratic gravity the sign of the coefficient of the Weyl squared term is uniquely fixed by the requirement that the massive pole is not a tachyon. The two possible signs correspond to tachyonic excitation or a massive unstable ghost (the Merlin resonance). We know of no way to make the tachyonic branch into a physically viable theory. However the unstable ghost branch leads to a unitary theory, albeit one with causality violation. We can see how this choice of sign influences physical observables by considering scattering of regular particles in the spin-two channel in which case the s-channel scattering goes through the Merlin resonance. The causality violation by the Merlin mode both dictates the sign of the higher order operator and also leads to a {\em negative} Wigner time advance in the scattering amplitude.

In potential scattering, Eisenbud and Wigner derived an implication of causality in the scattering phase
$\delta$ from a potential which vanishes identically outside of a radius $R$~\cite{Eisenbud,Wigner}. The time delay $\delta t$ is proportional to the energy derivative of the phase shift and is bounded
\beq
\delta t = 2 \frac{\partial \delta}{\partial E} \ge - \frac{2 m R}{p} \ \ .
\eeq
This can be found heuristically by Taylor expanding the outgoing wave phase shift
\beq
e^{i(kr - Et)} e^{2i\delta(E)}
\eeq
in powers of the energy and identifying the shift $e^{-iE(t-2\frac{\partial \delta}{\partial E})}$ as a time delay. The assumption that the potential vanishes outside of a radius $R$ is needed because the incoming wave can start scattering before reaching the center of the potential. For relativistic gravitational scattering the potential formally has infinite range, but handwaving estimates based on the uncertainty principle can be made that the right hand side of this equation should be no more than the order of the inverse of the center-of-mass energy. In practice applications are generally made by neglecting the right hand side of the equation, such that the time delay should be positive. Explicit calculations with realistic causal interactions support this. This is the reason that the phase shift advances in a counter-clockwise fashion on the Argand diagram.

However in gravitation scattering in the spin two channel, the phase motion is in the opposite direction. In the large N limit the spin two scattering proceeds through the direct s-channel graviton propagator\footnote{The invocation of the large N limit is useful because there is also a t-channel scattering. However the s-channel behaves as described here even away from the large N limit.}. This was originally analysed by Han and Willenbrock \cite{Han} independent of any UV completion - see also \cite{Aydemir:2012nz}. We have explored the reaction in more detail in the context of quadratic gravity. The basic point is that unitarity is maintained by the imaginary parts which arise in the graviton vacuum polarization diagram. However, those imaginary parts lead the phase to proceed clockwise on the Argand diagram. Within quadratic gravity, this motion is completed by a full resonance - the Merlin resonance - with clockwise motion of the phase past $-90^o$.

As detailed in the previous papers~\cite{Han, Aydemir:2012nz}, one finds that the partial wave amplitude $T_2$ has the form:
\beq\label{t2amplitude}
T_2(s) = - \frac{N_{\textrm{eff}}\kappa^2 s}{640 \pi}\,\bar{D}(s).
\eeq
where
\beq
\bar{D}^{-1}(s) =\, \left\{1- \frac{{\kappa}^2 s}{2\xi^2(\mu)}
- \frac{{\kappa}^2 s N_{\textrm{eff}}}{640\pi^2} \ln \left(\frac{s}{\mu^2}\right)
+ \frac{i {\kappa}^2 s N_{\textrm{eff}}}{640\pi}\right\}
\label{prop}
\eeq
Elastic unitarity requires $\textrm{Im} T_2 = |T_2|^2$. This is satisfied for an amplitude of the form
\beq
T_2(s) = \frac{A(s)}{f(s) -i A(s)}= \frac{A(s)[f(s)+i A(s)]}{f^2(s)+A^2(s)}
\eeq
for any real functions $f(s),~A(s)$. Since the imaginary part in the denominator comes from the logarithm, the unitarity condition is a relation between the tree-level scattering amplitude which determines the $A(s)$ in the numerator and the logarithm in the vacuum polarization which determines the imaginary part in the denominator. This is satisfied with
\beq
A(s) =  - \frac{N_{\textrm{eff}}\kappa^2 s}{640 \pi}.
\eeq
There is a correlation between the unusual sign of the imaginary part in the propagator and the sign of the scattering amplitude. This combination allows unitarity to be satisfied.

\begin{figure}[htb]
\begin{center}
\includegraphics[height=80mm,width=80mm]{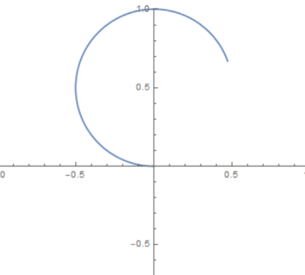}
\caption{Argand diagram for the phase shift in the J=2 channel. }
\label{phaseshift}
\end{center}
\end{figure}

\begin{figure}[htb]
\begin{center}
\includegraphics[height=80mm,width=80mm]{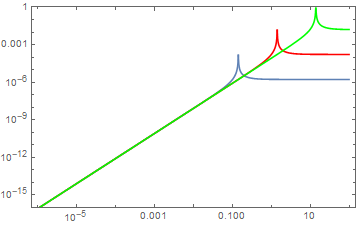}
\caption{Absolute value of the scattering amplitude in the J=2 channel. The curves from bottom to top correspond to $\xi^2= 0.1,~ 1.0,~10$.  }
\label{amp}
\end{center}
\end{figure}

The phase motion in this case is clockwise. This is seen in the Argand diagram of Figure \ref{phaseshift}, which has been continued through the resonance. In practice, the phase remains very small until one approaches close to the resonance. This low energy portion of this phase is universal. One can see this by considering the amplitude itself, for which the absolute value is shown in Figure \ref{amp} for different values of the parameter $\xi$. One sees that the amplitude is very small in the universal region - note the logarithmic scale. The phase motion happens when near the narrow resonance and this corresponds to the UV completion of the amplitude in quadratic gravity. For small values of $\xi$, this UV completion occurs below the Planck scale. However, for larger values of $\xi$ the amplitude can grow large before the UV completion happens. Of course, in other theories of gravity, the behaviour will resolve in different ways.

The main point here is that this phase behavior corresponds to a time advance rather than time delay. When this goes through a resonance, this is the defining characteristic of the Merlin mode, which propagates backwards in time.


\section{Fluctuations in the causal structure}

The previous discussions show us that one must be careful in using the familiar classical notion of causality in quantum settings. In particular, the classical picture of lightcones might not be a useful concept in quantum gravity, even in the effective field theory regime. But does this mean that one could, in principle, detect causality violations at low energy?

Here we wish to address this point. We want to show that, despite the fact that the usefulness of lightcones might be open to question in quantum gravity, we do not expect to verify unusual causal behavior in the low-energy regime. For such, let us get back to the discussion regarding the signs of coefficients of terms in an effective Lagrangian. This amounts to considering positivity constraints and effective lightcones~\cite{Adams:06}. We will discuss the consequences for the propagation of massless particles. For simplicity, assume that the gravitons are coupled with a massless scalar field $\pi$ with leading interaction
\beq
- \frac{\kappa}{2} h_{\mu\nu} T^{\mu\nu}
\eeq
where $T^{\mu\nu}$ is the energy-momentum tensor associated with the scalar field:
\beq
T_{\mu\nu} = \partial_{\mu} \pi \partial_{\nu} \pi
- \frac{1}{2} \eta_{\mu\nu} \partial_{\kappa} \pi \partial^{\kappa} \pi .
\eeq
By integrating out the gravitons, one finds the following contribution to the effective action
\beq
S_{\textrm{eff}} = \frac{1}{2} \int d^4 x \, \partial_{\mu}\pi(x) \partial^{\mu} \pi(x)
- \frac{\kappa^2}{8} \int d^4 x \int d^4 x' \, T^{\mu\nu}(x) D_{\mu\nu\alpha\beta}(x-x') T^{\alpha\beta}(x')
\eeq
where $D_{\mu\nu\alpha\beta}(x')$ is the propagator associated with the field $h_{\mu\nu}$. Using a de Donder gauge, the propagator can be brought to the form~\cite{unitarity,bos}
\bea
D_{\mu\nu\alpha\beta}(x-x') &=& \frac{1}{2} \Bigl( \eta_{\mu\alpha} \eta_{\nu\beta}
+ \eta_{\mu\beta} \eta_{\nu\alpha} - \eta_{\mu\nu} \eta_{\alpha\beta} \Bigr)
\int \frac{d^{4} q}{(2\pi)^4} e^{-i q \cdot (x-x')} D_2(q)
\nn\\
&=& \frac{1}{2} \Bigl( \eta_{\mu\alpha} \eta_{\nu\beta}
+ \eta_{\mu\beta} \eta_{\nu\alpha} - \eta_{\mu\nu} \eta_{\alpha\beta} \Bigr) D_{2}(x-x') .
\label{prop2}
\eea
This allows us to rewrite the effective action as
\beq
S_{\textrm{eff}} = \int d^4 x \, \biggl[ \frac{1}{2} \partial_{\mu}\pi(x) \partial^{\mu} \pi(x)
- \frac{\kappa^2}{16} \int d^4 x' \, D_{2}(x-x')
\Bigl( 2 T^{\mu\nu}(x) T_{\mu\nu}(x')
 - T^{\mu}_{\mu}(x) T^{\nu}_{\nu}(x') \Bigr)
 \biggr]
\eeq
or
\beq
S_{\textrm{eff}} = \int d^4 x \, \biggl[ \frac{1}{2} A_{\mu}(x) A^{\mu}(x)
- \frac{\kappa^2}{16} \int d^4 x' \, D_{2}(x-x')
\Bigl( 2 A^{\mu}(x) A^{\nu}(x) A_{\mu}(x') A_{\nu}(x')
 - A_{\lambda}(x) A^{\lambda}(x) A_{\kappa}(x') A^{\kappa}(x') \Bigr)
 \biggr]
\eeq
where $A_{\mu}(x) = \partial_{\mu} \pi(x)$. The effective Lagrangian is given by
\beq
{\cal L}_{\textrm{eff}} =  \frac{1}{2} A_{\mu}(x) A^{\mu}(x)
- \frac{\kappa^2}{16} \int d^4 x' \, D_{2}(x-x')
\Bigl( 2 A^{\mu}(x) A^{\nu}(x) A_{\mu}(x') A_{\nu}(x')
 - A_{\lambda}(x) A^{\lambda}(x) A_{\kappa}(x') A^{\kappa}(x') \Bigr).
\eeq
The equations of motion read
\beq
 \partial^{\mu} A_{\mu}(x)
- \frac{\kappa^2}{8} \partial^{x}_{\mu} \int d^4 z \, D_{2}(z)
\Bigl( 2 A^{\nu}(x) A^{\mu}(x-z) A_{\nu}(x-z)
- A^{\mu}(x) A_{\nu}(x-z) A^{\nu}(x-z)
\Bigr)
 = 0 .
\eeq
Now, let us expand our model around the solution $\partial_{\mu} \pi_0 = C_{\mu} \chi(x)$, where $C_{\mu}$ is a constant vector and $\chi(x)$ is a real scalar function. The linearized equations of motion for small fluctuations $\varphi = \pi - \pi_0$ around this background implies that
$A^{\mu} = \partial^{\mu}\varphi + C^{\mu} \chi(x)$. The equations of motion read
\beq\label{twoterms}
\biggl( \eta^{\mu\nu}
- \frac{\kappa^2}{4} \int d^4 z \, D_{2}(z) F(x-z) C^{\mu} C^{\nu} \biggr)
\partial_{\mu} \partial_{\nu}\varphi
- \frac{\kappa^2}{4} \int d^4 z \, D_{2}(z) \partial^{x}_{\mu} F(x-z) C^{\mu} C^{\nu}
 \partial_{\nu}\varphi + \cdots
= 0
\eeq
where $F(x) = \chi^2(x)$. We kept only the leading contribution linear in $\varphi$ and we considered that
$\kappa^2 ( C^{\nu} C_{\nu}) \ll 1$ so that all such terms can be neglected. In the low-energy regime, one can always argue that $\chi(x)$ is a slowly-varying function in such a way that all its derivatives can be dropped (including a ``source"  term linear in $\partial_{\mu} \chi$ which is of zeroth order in $\varphi$). Therefore
\beq
\biggl( \eta^{\mu\nu}
+ c_3(x) C^{\mu} C^{\nu} \biggr)
\partial_{\mu} \partial_{\nu}\varphi
+ \cdots
= 0 .
\eeq
where
\beq
c_3(x) \equiv - \frac{\kappa^2}{4} \int d^4 z \, D_{2}(z) F(x-z) .
\label{ident}
\eeq
An expansion in plane waves yields
\beq
p^{\mu} p_{\mu} + c_3(x) ( C \cdot p )^2 + \cdots = 0 .
\eeq
As discussed, in the low-energy regime we can drop all the derivatives of $\chi$ in which case $c_3$ is approximately constant. In this case we can explicitly identify a leading contribution in the above expansion -- which is given by the term multiplied by the coefficient $c_3$. Now we can write that
\beq
\int d^4 z \, D_{2}(z) F(x-z) \approx \int \frac{d^{4} q}{(2\pi)^4} D_{2}(q^2) F(q).
\eeq%
where we used the convolution theorem and a simple Taylor expansion. In this case, the function $F(q)$ can be envisaged as restricting the values of $q$ to the low-energy region. That means we should consider a low-energy representation for $D_{2}(q^2)$. At low energies, the leading contribution to the propagator is given by the usual form from general relativity. In Fourier space, and using a spectral representation, we see that
\beq
\int d^4 z \, D_{2}(z) F(x-z) < 0
\eeq
and $c_3$ is always positive. This implies the absence of superluminal excitations in the sense of Ref.~\cite{Adams:06}.

One could expect to find a remainder coming from Merlin modes. In the narrow-width approximation, the Merlin contribution to the propagator behaves as
\beq
iD_2 (q) = \frac{-i}{q^2-M^2 -i M \Gamma}
\eeq
where $\Gamma$ is the Merlin width. At low energy, this will produce the following contribution in the limit
$\Gamma \to 0$
\beq
iD_2 (z) \sim \frac{i \delta^{(4)}(z)}{M^2}
\eeq
and hence
\beq
\int d^4 z \, D_{2}(z) F(x-z) \approx \frac{F(0)}{M^2} > 0 .
\eeq
This show that the Merlin modes produce a negative shift in $c_3$ at low energies. However, since $M^2$ is of order of the Planck energy scale, this will produce a subleading effect which is virtually unobservable in this energy regime.

The linearized equation for the fluctuations $\varphi$ suggests a natural effective ``metric" (at least in the geometric optics limit
\beq
G_{\mu\nu} = \eta_{\mu\nu} + \frac{\kappa^2}{4} \int d^4 z \, D_{2}(z) F(x-z) C_{\mu} C_{\nu} + \cdots
\eeq
with which one could define effective lightcones and time evolution. However, this is only justified at low energies.

Unfortunately at high energies and for cases where $\chi(x)$ is not slowly varying, the discussion is more involved and a firm conclusion cannot be given. One could be tempted to employ naively a high-energy representation for $D_{2}(z)$ -- similar analysis as the one carried out above would reveal that, close to the Merlin resonance, $c_3$ would change sign. However the analysis itself is no longer valid. The first thing is the above linearization procedure: Although it works reasonably well at low energies, it is not clear whether this is a valid procedure at high energies -- one is not justified in dropping all derivatives of $\chi(x)$ in this regime. Moreover, the second term in Eq. \ref{twoterms} needs to be included and this invalidates the analysis. In addition $c_3(x)$ would have strong spacetime dependence. These features would seem to reinforce the uncertainty in the causal properties at small scales, of order of the Merlin width.

Despite such difficulties, our brief exploration does teach us two important lessons. The first is that, even though lightcones might not be a good concept in quantum gravity even in the long-wavelength case, at low energies the theory does not display explicit causality violation. The second lesson is that, although there are certainly other contributions at high energies whose effects should be analyzed properly, there is likely at least a certain level of causal uncertainty when close to the Planck scale. In any case, a proper treatment should investigate more deeply what happens to microcausality at high energies and this requires a UV complete theory. In principle, we know how to answer this question in quadratic gravity -- here the $S$-matrix is not analytic due to the Merlin resonance. An interesting course of action is to exploit the repercussion of this non-analyticity feature of scattering amplitudes upon the Pauli-Jordan function. And this is the topic of the following section.

\section{Causal uncertainty in quadratic gravity}

We now study with more detail the emergence of causality uncertainties in the specific context of quadratic gravity. As well known, in a causality preserving quantum field theory, the Pauli-Jordan function has support on the lightcone. This is usually refer to as the requirement of microcausality: Any two local observables at spacelike separation must commute~\cite{GellMann:1954db}. We will see how this can be violated in quadratic gravity.

Let us begin our study by considering the retarded and advanced Green's functions. Usually such functions vanish for spacelike separations. Taking into account the one-loop vacuum polarization, the spin-$2$ part of the retarded Green's function is given by~\cite{Donoghue:2018izj,unitarity}
\bea
i D^{\textrm{ret}}_{\mu\nu\alpha\beta}(x-x') &=& \kappa^2
\widetilde{{\cal P}}^{(2)}_{\mu\nu\alpha\beta}(\partial_{x})
\int \frac{d^{4} q}{(2\pi)^4} e^{-i q \cdot (x-x')} i D^{\textrm{ret}}_2(q)
\nn\\
\widetilde{{\cal P}}^{(2)}_{\mu\nu\rho\sigma}(\partial) &=& \frac{1}{2}(\theta_{\mu\rho}\theta_{\nu\sigma}
+ \theta_{\mu\sigma}\theta_{\nu\rho}) - \frac{1}{3} \theta_{\mu\nu}\theta_{\rho\sigma},
\,\,\,\,
\theta_{\mu\nu} = \eta_{\mu\nu} - \frac{\partial_{\mu}\partial_{\nu}}{\Box^2}
\nn\\
D^{\textrm{ret}}_2(q) &=& \frac{1}{( q^{0} + i \epsilon )^{2} - {\bf q}^2}
- \frac{1}{q^{2} - m^{2}_{r} - i \gamma \theta(q^2) \bigl( \theta(q^{0}) - \theta(- q^{0}) \bigr)}
\nn\\
m^{2}_{r} &=& \frac{2 \xi^2(m^{2}_{r})}{\kappa^2}
+ \frac{\xi^2(m^{2}_{r}) N_{\textrm{eff}} }{320 \pi^2 \kappa^2} \ln\left( \frac{|q^2|}{m^{2}_{r}} \right)
\approx \frac{2 \xi^2( m^{2}_{r} )}{\kappa^2}
\nn\\
\gamma &\approx& m^{2}_{r} \frac{ \xi^2(m^{2}_{r}) N_{\textrm{eff}} }{320 \pi}
\eea
where we consider that, in the weak-coupling limit $\xi << 1$, $\gamma$ is a small number so that one may drop the term $( \xi^2 N_{\textrm{eff}} / 320 \pi )^2$ and the logarithmic term. Contour integration will lead us to the result
\bea
D^{\textrm{ret}}_{\mu\nu\alpha\beta}(x) &=& \kappa^2
\widetilde{{\cal P}}^{(2)}_{\mu\nu\alpha\beta}(\partial_{x})
\int \frac{d^{3} q}{(2\pi)^3} \Biggl\{ - i \theta(t)
\left[ \frac{e^{-i ( \omega_{q} t - {\bf q} \cdot {\bf x}) } }{2 \omega_{q}}
- \frac{e^{i ( \omega_{q} t - {\bf q} \cdot {\bf x}) } }{2 \omega_{q}} \right]
\nn\\
&+&  i \theta(-t) \left[ \frac{e^{-i ( E_{q} t - {\bf q} \cdot {\bf x}) } }{2 \left( E_{q} + i \frac{\gamma}{2 E_{q}} \right) }
- \frac{e^{i ( E_{q} t - {\bf q} \cdot {\bf x}) } }{2 \left( E_{q} - i \frac{\gamma}{2 E_{q} } \right)} \right]
e^{- \frac{\gamma |t|}{2 E_{q}}}
\Biggr\}
\eea
where $\omega_{q} = |{\bf q}|$ and $E_{q} = \sqrt{ {\bf q}^2 + m^{2}_{r} }$ . So we see that not only there is an unusual term for $t<0$ but also such a term fails to vanish for spacelike separations (the easiest way to see this is to consider a frame in which the separation is purely spatial, that is, $t=0$). In turn, for the advanced Green's function, a similar reasoning lead us to the following result:
\bea
D^{\textrm{adv}}_{\mu\nu\alpha\beta}(x) &=& \kappa^2
\widetilde{{\cal P}}^{(2)}_{\mu\nu\alpha\beta}(\partial_{x})
\int \frac{d^{3} q}{(2\pi)^3} \Biggl\{ i \theta(-t)
\left[ \frac{e^{-i ( \omega_{q} t - {\bf q} \cdot {\bf x}) } }{2 \omega_{q}}
- \frac{e^{i ( \omega_{q} t - {\bf q} \cdot {\bf x}) } }{2 \omega_{q}} \right]
\nn\\
&-&  i \theta(t) \left[ \frac{e^{-i ( E_{q} t - {\bf q} \cdot {\bf x}) } }{2 \left( E_{q} - i \frac{\gamma}{2 E_{q}} \right) }
- \frac{e^{i ( E_{q} t - {\bf q} \cdot {\bf x}) } }{2 \left( E_{q} + i \frac{\gamma}{2 E_{q} } \right)} \right]
e^{- \frac{\gamma t}{2 E_{q}}}
\Biggr\} .
\eea
Hence one verifies the presence of an unusual term for $t>0$. In addition, the advanced Green's function also fails to vanish for spacelike separations.

To carefully implement a study on causality, one must duly calculate the Pauli-Jordan function $G_{\mu\nu\alpha\beta}(x,x') = [h_{\mu\nu}(x),h_{\alpha\beta}(x')]$. Therefore, let us consider the associated Wightman functions. Again we investigate the simple case of a massless scalar field coupled with gravity. The full gravitational Pauli-Jordan function is given by
\beq
G_{\mu\nu\alpha\beta}(x,x') = G_{\mu\nu\alpha\beta}^{+}(x,x') - G_{\mu\nu\alpha\beta}^{-}(x,x')
\label{PJ}
\eeq
where $G_{\mu\nu\alpha\beta}^{\pm}$ are Wightman functions. From the results discussed at length in the Appendix, we find that
\beq
G_{\mu\nu\alpha\beta}(x,x') = G^{0}_{\mu\nu\alpha\beta}(x-x')
- \int d^4 x'' D^{0\textrm{ret}}_{\gamma\delta\alpha\beta}(x-x'')
F^{\gamma\delta}_{\ \ \mu\nu}(x'',x')
\eeq
where $G^{0}_{\mu\nu\alpha\beta}(x-x')$ is the unperturbed Pauli-Jordan function and we have defined
\beq
F^{\gamma\delta}_{\ \ \mu\nu}(x'',x') =
\biggl\langle \frac{\delta {\cal L}^{+}_{\textrm{int}} }{\delta h^{+}_{\gamma\delta}(x'')} [\Phi^{+}(x'')]
\, h^{-}_{\mu\nu}(x') \biggr\rangle \Bigg|_{\Phi^{-} = \Phi^{+} = \Phi}
+ \biggl\langle
h^{+}_{\mu\nu}(x') \,
\frac{\delta {\cal L}^{-}_{\textrm{int}} }{\delta h^{-}_{\gamma\delta}(x'')} [\Phi^{-}(x'')]
\biggr\rangle \Bigg|_{\Phi^{-} = \Phi^{+} = \Phi}  .
\eeq
In this equation, $D^{0\textrm{ret}}_{\gamma\delta\alpha\beta}(x-x')$ is the unperturbed retarded Green's function and ${\cal L}_{\textrm{int}}$ describes gravitational interactions. We use the in-in formalism, so that fields and sources are doubled (which clarifies the $\pm$ appearing above) and $\Phi_{\pm}$ denotes the collection of the fields. For a thorough discussion of terminology and notations, see the Appendix.

For cubic interactions, $F^{\gamma\delta}_{\ \ \mu\nu}$ is simply the symmetric correlation function
\beq
F^{\gamma\delta}_{\ \ \mu\nu}(x'',x') = - \frac{\kappa}{2}
\bigl\langle \bigl\{ T^{\gamma\delta}(x''), h_{\mu\nu}(x') \bigr\} \bigr\rangle
\eeq
where $T^{\gamma\delta}$ is the energy-momentum tensor associated with the field $\Phi$. Using standard properties of the Heaviside theta function, we can also write that
\bea
F^{\gamma\delta}_{\ \ \mu\nu}(x'',x') &=& - \frac{\kappa}{2}
\left[ \bigl\langle T \bigl\{ h_{\mu\nu}(x') T^{\gamma\delta}(x'') \bigr\} \bigr\rangle
+ \bigl\langle \bar{T} \bigl\{ T^{\gamma\delta}(x'') h_{\mu\nu}(x') \bigr\} \bigr\rangle \right]
\nn\\
&=& - \frac{\kappa}{2} \lim_{x''' \to x''} {\cal D}^{\gamma\delta}(x'',x''')
\left[ \bigl\langle T \bigl\{ h_{\mu\nu}(x') \Phi(x'')\Phi(x''') \bigr\} \bigr\rangle
+ \bigl\langle \bar{T} \bigl\{ \Phi(x'')\Phi(x''') h_{\mu\nu}(x') \bigr\} \bigr\rangle
+ \cdots \right]
\label{45}
\eea
with $\bar{T}$ being the anti-time-ordering operation and
${\cal D}^{\gamma\delta}(x'',x''')$ is a non-local differential operator obtained from the definition of $T^{\gamma\delta}$. For instance, for a massless scalar field, it has the form
\beq
{\cal D}_{\mu\nu}(x',x'') = \frac{1}{2} \Bigl( \partial_{\mu\prime} \partial_{\nu\prime\prime}
+ \partial_{\mu\prime\prime} \partial_{\nu\prime} \Bigr)
- \frac{1}{2} \eta_{\mu\nu} \partial_{\kappa\prime} \partial^{\kappa\prime\prime}
\eeq
where the prime (or the double prime) on a derivative indicates that it acts at $x^{\prime}$ (or $x^{\prime\prime}$). The ellipsis denotes contact interactions that appear when we perform derivatives of theta functions.  Moreover, one must consider either $t' > t'', t'''$ or $t' < t'', t'''$ in order to make sense to move from the first line to the second line of Eq.~(\ref{45}). What is left one can identify with possible contributions to S-matrix elements (with a different time direction for the second term). So, up to contact interactions, we are able to express $F^{\gamma\delta}_{\ \ \mu\nu}$ in terms of derivatives of three point functions $\langle h \Phi \Phi \rangle$, which have known expressions. For instance, at one loop, one would find the following typical terms coming from triangle diagrams involving gravitons (or Merlins) and scalars
$\phi$ for the time-ordered contribution~\footnote{For the case of Merlin particles flowing through an internal line of a given Feynman diagram, these only have the form of one-loop diagrams -- a Merlin particle is unstable and hence we should use a resummed form for its propagator. This is a consequence of the fact that ordinary perturbation theory breaks down in the resonance region of a diagram with an unstable internal propagator and, in order to solve this issue, the unstable propagator should be carefully resummed.}:
\beq
\bigl\langle T \bigl\{ h_{\mu\nu}(x') \phi(x'')\phi(x''') \bigr\} \bigr\rangle^{(1a)}
\, = \, \includegraphics[height=2.7cm,valign=c]{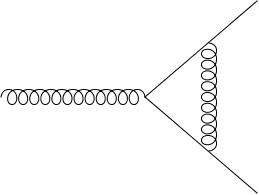}
\eeq
and
\beq
\bigl\langle T \bigl\{ h_{\mu\nu}(x') \phi(x'')\phi(x''') \bigr\} \bigr\rangle^{(1b)}
\, = \, \includegraphics[height=3.5cm,valign=c]{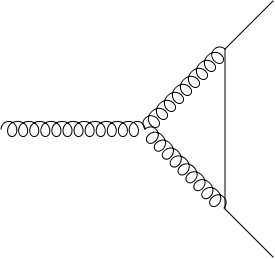}
\eeq
where curly lines denote graviton/Merlin propagators and solid lines represent scalar propagators. There are also terms associated with the gravitational three point function, as well as $3$-point functions involving the ghost-graviton (or ghost-Merlin) vertex. Similar terms can be found for the anti-time-ordered contribution. Consider now the narrow-width approximation. As discussed above, Merlin particles carry an opposite arrow of causality in comparison with normal unstable particles, so factors of $i \gamma$ associated with the width of the Merlin particle change sign in the corresponding propagators (which also acquire an overall minus sign). This implies that, when discussing the associated scattering amplitudes, one would find poles above the branch cut across the real axis when Merlin modes appear in loops, rendering $S$-matrix elements (involving, e.g., scattering of graviton and scalar particles) to be non analytic in the physical region. This non-analyticity feature comes exactly from the above diagrams containing Merlin propagators inside the loops, which are also present in $F^{\gamma\delta}_{\ \ \mu\nu}$.

As we have shown above, the retarded Green's function, in addition to the usual term, contains an anomalous term associated with propagation backward in time. Since
$D^{0\textrm{ret}}_{\gamma\delta\alpha\beta}$ is an unperturbed Green's function, the spin-$2$ contribution must have the form
\bea
D^{\textrm{ret}}_{\mu\nu\alpha\beta}(x-x') &=& \kappa^2
\widetilde{{\cal P}}^{(2)}_{\mu\nu\alpha\beta}(\partial_{x})
\Biggl[ - \frac{1}{4\pi |{\bf x} - {\bf x}'|} \delta\Bigl( t' - (t-|{\bf x} - {\bf x}'|) \Bigr)
+ \frac{1}{4\pi |{\bf x} - {\bf x}'|} \delta\Bigl( t' - (t+|{\bf x} - {\bf x}'|) \Bigr)
\nn\\
&-& \theta(t-t') \theta(\sigma_{0})
\frac{m_0 J_{1}\Bigl(m_0 \sqrt{\sigma_{0}}\Bigr)}{4\pi \sqrt{\sigma_{0}}} \Biggr]
\eea
where $m_0^2 = 2 \xi^2 /\kappa^2$. For our purposes the term with the Bessel function $J_{1}$ will not be important so we will not study its contribution in detail. The Pauli-Jordan function reads
\bea
G_{\mu\nu\alpha\beta}(x,x') &=& G^{0}_{\mu\nu\alpha\beta}(x-x')
-  \frac{\kappa^2}{4\pi} \widetilde{{\cal P}}^{(2)}_{\gamma\delta\alpha\beta}(\partial_{x})
 \int \frac{d^3x''}{|{\bf x} - {\bf x}''|} \left[ - F^{\gamma\delta}_{\ \ \mu\nu}(t''-t'; {\bf x}'' - {\bf x}')
\bigg|_{t''= t-|{\bf x} - {\bf x}''|}
\right.
\nn\\
&+& \left. F^{\gamma\delta}_{\ \ \mu\nu}(t''-t'; {\bf x}'' - {\bf x}') \bigg|_{t''= t+|{\bf x} - {\bf x}''|} + \cdots \right]
\eea
where the ellipsis correspond to the contribution coming from the Bessel function $J_{1}$. Now consider that ${\bf x} = - {\bf x'} = r {\bf \hat{n}}$. Furthermore, consider the Pauli-Jordan function for large values of  $r$. Then one has that
$$
|{\bf x} - {\bf x}''| \approx r - {\bf x}'' \cdot {\bf \hat{n}}
$$
Introducing the Fourier transform
$$
F^{\gamma\delta}_{\ \ \mu\nu}(x) = \int \frac{d^4 p}{(2\pi)^4} F^{\gamma\delta}_{\ \ \mu\nu}(p)
e^{- i p \cdot x}
$$
one obtains
\bea
G_{\mu\nu\alpha\beta}(x,x') &\approx& G^{0}_{\mu\nu\alpha\beta}(x-x')
-  \frac{\kappa^2}{2\pi} \widetilde{{\cal P}}^{(2)}_{\gamma\delta\alpha\beta}(\partial_{x})
\int  \frac{d^3x''}{|{\bf x} - {\bf x}'|} \int \frac{d^4 p}{(2\pi)^4} F^{\gamma\delta}_{\ \ \mu\nu}(p)
\left[ - e^{- i p^{0} (t-t'-r+{\bf x}'' \cdot {\bf \hat{n}})} e^{i {\bf p} \cdot ({\bf x}'' - {\bf x}') }
\right.
\nn\\
&+& \left. e^{- i p^{0} (t-t'+r-{\bf x}'' \cdot {\bf \hat{n}})} e^{i {\bf p} \cdot ({\bf x}'' - {\bf x}') } + \cdots \right]
\nn\\
&=& G^{0}_{\mu\nu\alpha\beta}(x-x')
+  \frac{\kappa^2}{2\pi} \widetilde{{\cal P}}^{(2)}_{\gamma\delta\alpha\beta}(\partial_{x})
\left\{ \frac{1}{|{\bf x} - {\bf x}'|}  \int \frac{d p^{0}}{2\pi}
\left[ F^{\gamma\delta}_{\ \ \mu\nu}(p^0, p^{0} {\bf \hat{n}} ) \, e^{i p^{0} z}
- F^{\gamma\delta}_{\ \ \mu\nu}(p^0, - p^{0} {\bf \hat{n}} ) \, e^{- i p^{0} z'}
+ \cdots \right] \right\}
\nn\\
\eea
where $z =  |{\bf x} - {\bf x}'| - (t-t')$ and $z' =  |{\bf x} - {\bf x}'| + (t-t')$. Let us assume that $t-t'>0$ and $z>0$. Then the first term is the most interesting one. One may evaluate the first term using contour-integration methods. Consider an appropriate contour above the real axis. If the integral gives a zero value, that means that the integrand is an analytic function in the upper half $p^0$ plane. As a consequence, the Pauli-Jordan function will vanish for a spacelike separation $z>0$. However, as extensively discussed above, $F^{\gamma\delta}_{\ \ \mu\nu}(p)$ is in general not an analytic function in the upper half $p^0$ plane due to the presence of poles coming from the existence of Merlin modes. Hence the full Pauli-Jordan function fails to vanish for spacelike separations in quadratic gravity.

The other extreme from quadratic gravity could be the theory of causal sets \cite{Bombelli:1987aa, Surya:2019ndm}. Here the theory is described by configurations which exhibit definite causal ordering. However, even here where true causality is built into the foundation of the theory, there can be causal uncertainties. The causal sets can appear like a smoothed spacetime on scales much larger than the discreteness scale. However, the quantum theory will involve fluctuations among the causal sets, with different light paths and times. The result is a fluctuating uncertainty of macroscopic causality even though the underlying theory is defined by causal relations.


\section{Summary}

Because the size of quantum gravitational effects grows with energy, metric fluctuations on short scales are expected to be large, if the metric even survives as a dynamical variable. On these scales, standard ideas of causality likely no longer make sense. While we cannot treat the deep quantum regime with reliability, we can see some causality uncertainty arising at lower energy where the calculations are more meaningful. If the metric is not the relevant dynamical variable at high energies, the situation is likely even more uncertain as the spacetime intervals become ill-defined.

We have shown here through the discussion of many examples that causal uncertainty in field theory should be expected in quantum gravity. This emerges even at low energies, as the analysis of the effective field theory of gravity clearly shows that the concept of a lightcone becomes problematic. We have also discussed this issue in the context of a particular UV completion, that of quadratic gravity, where two possible causal flows are present at energies close to the Planck scale. In this case it is clear how causality in a traditional sense should be envisaged as an emergent macroscopic phenomenon.

One traditional way of thinking on causal uncertainty concerns the analyticity of the S-matrix. Namely, when amplitudes fail to satisfy the standard analyticity axioms of S-matrix theory, one verifies some sort of causality violation. This is exactly what happens in Lee-Wick-like theories, where the poles associated with Merlin modes are located on the physical sheet. Here we have shown how this impacts the calculation of the Pauli-Jordan function, which embodies the concept of microcausality in field theory. In our evaluation, we have demonstrated that the full Pauli-Jordan function fails to vanish for spacelike separations in quadratic gravity.

One other possible way to study causal uncertainty is through an analysis of the Shapiro's time delay. Indeed, higher derivative corrections to the graviton three-point function lead to a potential causality violation unless we get contributions from an infinite tower of extra massive particles with higher spins, which implies a Regge behavior for the amplitudes -- this is verified in a weakly coupled string theory~\cite{Camanho:2016,DAppollonio:2015}. This also happens in large $N$ QCD coupled to gravity~\cite{Kaplan:2020}; in this case there is also an alternative (but similar) resolution: The higher-spin states well below the Planck scale can come from the QCD sector. All such examples displaying causality violations (and the associated resolutions) come from higher-derivative corrections to gravitational interactions. The situation in quadratic gravity is somewhat different; if, on one hand, $3$-particle amplitudes involving only physical gravitons do not display contributions coming from higher-order derivative terms (and this result generalizes to an arbitrary number of external gravitons)~\cite{Johansson:18,DM:21}, on the other hand the existence of causal uncertainty comes from the fact that there are two sets of modes that follow different causal directions; that is, if one takes such sets in separation, there should be no issues with causality~\cite{Donoghue:2019ecz,Donoghue:2020mdd}. The problem lies when one takes them together -- ``dueling arrows of causality" near the Planck scale. All associated amplitudes will present normal behavior, which is reflected by the fact that the theory is unitary~\cite{unitarity}. Moreover, we expect Shapiro's time delay to exhibit such features. So a potential ``time advance" instead of time delay could be potentially verified when one takes a theory containing normal modes together with Merlin modes. The full analysis of the Shapiro's time delay in quadratic gravity would be interesting to explore, and we hope to do this in the near future.

The uncertainty in standard ideas of casuality can also be seen in gravitational models where there is a causal order, such as causal sets \cite{Cohen:2019qgh}. The reasoning is similar in that there are quantum fluctuations in the elements which blurs the ordering on the average. It would be interesting to see if similar uncertainties could be quantified in highly quantum numerical simulations in schemes such as Causal Dynamical Triangulations \cite{Loll:2019rdj}.

Many discussions of quantum gravity continue to use classical concepts such as lightcones and background metrics. These may be permissable in regimes where the quantum effects are tiny. But we are often interested in regimes where quantum effects are important, and then these classical ideas are likely illegitimate. Many discussions of quantum gravity continue to make use of classical notions, such a Penrose diagrams. These are not well defined concepts in quantum gravity. Moreover, to the extent that these effects could depend on the UV completion, this provides an uncertainty of unknown magnitude - quite possibly large - to our discussions. The causality uncertainty which we have explored is one of the newer quantum features which needs to be dealt with.  We need to develop ways of understanding the deep quantum regime which acknowledges the changes which occur in some of our standard expectations.

\section*{Acknowledgements} We thank Sumati Surya for discussions about causal uncertainties. The work of JFD has been partially supported by the US National Science Foundation under grant NSF-PHY18-20675. The work of GM has been partially supported by  Conselho Nacional de Desenvolvimento Cient\'ifico e Tecnol\'ogico - CNPq under grant 307578/2015-1 (GM) and Funda\c{c}\~ao Carlos Chagas Filho de Amparo \`a Pesquisa do Estado do Rio de Janeiro - FAPERJ under grant E-26/202.725/2018.

\appendix

\section{Appendix: Derivation of the Wightman functions in the in-in formalism}
\label{app1}

In this Appendix we discuss in detail the Wightman functions in the in-in formalism. We also shall derive the necessary Schwinger-Dyson equations. For simplicity, consider a coupling with a massless scalar field:
\beq
S = \int d^{4} x \left[ - \frac{1}{2} h^{\mu\nu} \Box_{\mu\nu\alpha\beta} h^{\alpha\beta}
- \frac{1}{2} \phi \Box \phi + \bar{C}_{\alpha} M^{\alpha}_{\ \beta} C^{\beta}
+ {\cal L}_{\textrm{int}}[h,\phi,C,\bar{C}]  \right]
\eeq
where as above $\Box_{\mu\nu\alpha\beta}$ is a shorthand for the graviton kinetic terms (with the gauge-fixing contributions already included). In the above equation, $C^{\beta}$ is the standard Faddeev-Popov ghost field. After doubling the fields and sources, one obtains
\bea
Z[J^{\pm},K^{\pm},\sigma_{\pm},\bar{\sigma}^{\pm}]
&=& \left(  \det {\cal D}^{\mu\nu} \right)^{1/2}
\int D\Phi_{+} D\Phi_{-} \exp\Biggl\{i S[\Phi_{+}] - i S^*[\Phi_{-}] +
i \int d^{4}x \left[ J_{+}^{\mu\nu} h^{+}_{\mu\nu} + K^{+} \phi^{+} + \bar{\sigma}^{+}_{\alpha} C_{+}^{\alpha}
+ \bar{C}_{+}^{\alpha} \sigma^{+}_{\alpha} \right]
\nn\\
&-& i \int d^{4}x \left[ J_{-}^{\mu\nu} h^{-}_{\mu\nu} + K^{-} \phi^{-} + \bar{\sigma}^{-}_{\alpha} C_{-}^{\alpha}
+ \bar{C}_{-}^{\alpha} \sigma^{-}_{\alpha} \right]
 \Biggr\}
\eea
where $\Phi_{\pm}$ denotes the collection of the fields $h^{\pm}_{\mu\nu}, \phi^{\pm}, C_{\pm}^{\alpha},  \bar{C}_{\pm}^{\alpha}$ and $S^*$  indicates that in this functional $i \epsilon \to - i \epsilon$.
Gravitational Green's functions can be derived in the usual way, namely
\bea
&-& \frac{1}{Z[0]} \frac{1}{(-i)^{m+1}} \frac{1}{(i)^{n+1}} \frac{\delta^{m}}{ \delta J_{-}^{\mu_{1}\nu_{1}}(y_{1})
\cdots \delta J_{-}^{\mu_{m}\nu_{m}}(y_{m}) }
\frac{\delta^{n}}{ \delta J_{+}^{\alpha_{1}\beta_{1}}(x_{1})
\cdots \delta J_{+}^{\alpha_{n}\beta_{n}}(x_{n}) } Z[J^{\pm},K^{\pm},\sigma_{\pm},\bar{\sigma}^{\pm}]
\bigg|_{J^{\pm}=K^{\pm}=\sigma_{\pm}=\bar{\sigma}^{\pm}=0}
\nn\\
&=& \left\langle
T \Bigl\{ h^{+}_{\alpha_{1}\beta_{1}}(x_{1}) \cdots h^{+}_{\alpha_{n}\beta_{n}}(x_{n}) \Bigr\}
\bar{T} \Bigl\{ h^{-}_{\mu_{1}\nu_{1}}(y_{1}) \cdots h^{-}_{\mu_{m}\nu_{m}}(y_{m}) \Bigr\}
\right\rangle
\eea
where $\bar{T}$ denotes anti-time-ordering operation. In particular, for $n=m=1$ one finds the Wightman function
\beq
- \frac{1}{Z[0]} \frac{\delta}{ \delta J_{-}^{\mu\nu}(x)} \frac{\delta}{ \delta J_{+}^{\alpha\beta}(x')}
Z[J^{\pm},K^{\pm},\sigma_{\pm},\bar{\sigma}^{\pm}]
\bigg|_{J^{\pm}=K^{\pm}=\sigma_{\pm}=\bar{\sigma}^{\pm}=0}
= \langle h^{+}_{\mu\nu}(x') h^{-}_{\alpha\beta}(x) \rangle =  G_{\mu\nu\alpha\beta}^{-}(x,x').
\eeq
Likewise
\beq
- \frac{1}{Z[0]} \frac{\delta}{ \delta J_{-}^{\alpha\beta}(x')} \frac{\delta}{ \delta J_{+}^{\mu\nu}(x)}
Z[J^{\pm},K^{\pm},\sigma_{\pm},\bar{\sigma}^{\pm}]
\bigg|_{J^{\pm}=K^{\pm}=\sigma_{\pm}=\bar{\sigma}^{\pm}=0}
= \langle h^{+}_{\mu\nu}(x) h^{-}_{\alpha\beta}(x') \rangle = G_{\mu\nu\alpha\beta}^{+}(x,x') .
\eeq
Now let us derive the Schwinger-Dyson equation for the gravitational Wightman functions. We start with the Schwinger-Dyson differential equations, given by
\beq
\left\{ \frac{\delta S_{\pm}}{\delta h_{\pm}^{\mu\nu}}\left[ -i \frac{\delta}{\delta J^{\pm}},
- i \frac{\delta}{\delta K^{\pm}}, - i \frac{\delta}{\delta \bar{\sigma}^{\pm}}, i \frac{\delta}{\delta \sigma^{\pm}} \right] + J_{\mu\nu}^{\pm} \right\}
Z[J^{\pm},K^{\pm},\sigma_{\pm},\bar{\sigma}^{\pm}] = 0
\eeq
where $S_{\pm} = S[\Phi_{\pm}]$. Performing the functional derivative, one finds that
\beq
\left\{ i \Box_{\mu\nu\alpha\beta} \frac{\delta}{\delta J^{\pm}_{\alpha\beta}}
+ \frac{\delta {\cal L}^{\pm}_{\textrm{int}} }{\delta h_{\pm}^{\mu\nu}}\left[ -i \frac{\delta}{\delta J^{\pm}},
- i \frac{\delta}{\delta K^{\pm}}, - i \frac{\delta}{\delta \bar{\sigma}^{\pm}}, i \frac{\delta}{\delta \sigma^{\pm}} \right] + J_{\mu\nu}^{\pm} \right\}
Z[J^{\pm},K^{\pm},\sigma_{\pm},\bar{\sigma}^{\pm}] = 0 .
\eeq
Performing a further derivative with respect to $\delta / \delta J_{\mp}$, one finds that
\beq
 i \Box^{x}_{\mu\nu\alpha\beta} \frac{\delta^2 Z}{\delta J^{\mp}_{\gamma\delta}(x')
 \delta J^{\pm}_{\alpha\beta}(x)}
+ \frac{\delta {\cal L}^{\pm}_{\textrm{int}} }{\delta h_{\pm}^{\mu\nu}(x)}\left[ -i \frac{\delta}{\delta J^{\pm}},
- i \frac{\delta}{\delta K^{\pm}}, - i \frac{\delta}{\delta \bar{\sigma}^{\pm}}, i \frac{\delta}{\delta \sigma^{\pm}} \right] \frac{\delta Z}{\delta J^{\mp}_{\gamma\delta}(x')} = 0  .
\eeq
By taking further derivatives with respect to the external sources one can obtain the complete set of Schwinger-Dyson equations. In terms of correlation functions, one finds (setting all external sources to zero):
\bea
- i \Box^{x}_{\mu\nu\alpha\beta} G^{\gamma\delta\alpha\beta}_{+}(x,x')
-i \biggl\langle \frac{\delta {\cal L}^{+}_{\textrm{int}} }{\delta h_{+}^{\mu\nu}(x)} [\Phi^{+}(x)]
\, h^{\gamma\delta}_{-}(x') \biggr\rangle \Bigg|_{\Phi^{-} = \Phi^{+} = \Phi} &=& 0
\nn\\
- i \Box^{x}_{\mu\nu\alpha\beta} G^{\gamma\delta\alpha\beta}_{-}(x,x')
+ i \biggl\langle  h^{\gamma\delta}_{+}(x')
\frac{\delta {\cal L}^{-}_{\textrm{int}} }{\delta h_{-}^{\mu\nu}(x)} [\Phi^{-}(x)] \,
 \biggr\rangle \Bigg|_{\Phi^{-} = \Phi^{+} = \Phi} &=& 0 .
\eea
Given appropriate boundary conditions, the solutions of the above equations read
\bea
G^{+}_{\mu\nu\alpha\beta}(x,x')
&=& G^{0+}_{\mu\nu\alpha\beta}(x-x') - \int d^4 x'' D^{0\textrm{ret}}_{\gamma\delta\alpha\beta}(x-x'')
\biggl\langle \frac{\delta {\cal L}^{+}_{\textrm{int}} }{\delta h^{+}_{\gamma\delta}(x'')} [\Phi^{+}(x'')]
\, h^{-}_{\mu\nu}(x') \biggr\rangle \Bigg|_{\Phi^{-} = \Phi^{+} = \Phi}
\nn\\
G^{-}_{\mu\nu\alpha\beta}(x,x')
&=& G^{0-}_{\mu\nu\alpha\beta}(x-x') + \int d^4 x'' D^{0\textrm{ret}}_{\gamma\delta\alpha\beta}(x-x'')
\biggl\langle h^{+}_{\mu\nu}(x')  \,
\frac{\delta {\cal L}^{-}_{\textrm{int}} }{\delta h^{-}_{\gamma\delta}(x'')} [\Phi^{-}(x'')]
\biggr\rangle \Bigg|_{\Phi^{-} = \Phi^{+} = \Phi}
\eea
where $G^{0 \pm}_{\mu\nu\alpha\beta}(x-x')$ obeys the equation
\beq
\Box_{x}^{\mu\nu\alpha\beta} G^{0 \pm}_{\alpha\beta\gamma\delta}(x-x') = 0
\eeq
and $D^{0\textrm{ret}}_{\gamma\delta\alpha\beta}(x-x')$ is the unperturbed retarded Green's function which obeys
\beq
\Box_{x}^{\mu\nu\alpha\beta} D^{0\textrm{ret}}_{\alpha\beta\gamma\delta}(x-x')
= I^{\mu\nu}_{\ \ \gamma\delta} \delta(x-x')
\eeq
where (in a flat background)
\beq
I_{\mu\nu\alpha\beta} \equiv \frac{1}{2} \Bigl( \eta_{\mu\alpha} \eta_{\nu\beta}
+ \eta_{\mu\beta} \eta_{\nu\alpha} \Bigr) .
\eeq
In possession of Wightman's functions, one can explicitly construct an expression for the Pauli-Jordan function, given by Eq.~(\ref{PJ}).

 \end{document}